\documentclass[prb,
superscriptaddress,showpacs,amsmath,amssymb]{revtex4}

\begin{document}

\title{Thermodynamic and quantum fluctuations of horizon area}

\author{G.E.~Volovik}
\affiliation{Landau Institute for Theoretical Physics, acad. Semyonov av., 1a, 142432,
Chernogolovka, Russia}

\date{\today}

\begin{abstract}
 The event horizon is a source of irreversibility, analogous to statistical irreversibility. This is why for systems with an event horizon there is no difference between quantum and thermal fluctuations. Quantum processes of quantum tunneling determine the thermodynamics of these systems, their temperatures, entropies and fluctuations. We considered three examples of entropy variance that support this point of view: (i) the variance of the area of the black hole horizon, obtained by consideration of quantum fluctuations; (ii) the variance of the entropy of the Hubble volume in the de Sitter state, obtained by consideration of thermal fluctuations; and (iii) the variance of entropy in integers in the Planckon model, determined by the Poisson distribution.
 \end{abstract}
\pacs{
}

\maketitle

\tableofcontents

 \section{Introduction}
  \label{introduction}

We discuss the quantum and thermal fluctuations of the entropy related to the event horizons.
In Section \ref{quantum} the quantum fluctuations of the black hole area are considered in terms of the pair of the canonically conjugate gravitational variables. In Section \ref{Thermal} the thermal fluctuations of the entropy of the Hubble volume in the de Sitter state are considered. The variance of the thermal entropy exactly corresponds to the quantum fluctuations of the area of the horizon.
This demonstrates the connection between quantum and thermodynamic fluctuations in the systems with event horizon. In Section \ref{Planckons} the thermodynamic and quantum fluctuations of black hole are discussed in the toy model, which is suggested by the non-extensive Tsallis-Cirto statistics. 
It is ensemble of the correlated pairs of Planckons, where each Planckon is the object with reduced Planck mass and zero entropy.

 \section{Quantum fluctuations of black hole entropy}
  \label{quantum}
  
Here we consider quantum fluctuations of the area of the black hole horizon using the gravitational canonically conjugate variables. In the case of the black hole, these are the horizon area $A$ and the gravitational coupling $K=1/(4G)$. For the black hole with mass $M$ one has:
\begin{equation}
\frac{\partial A}{\partial \tau}=\frac{\partial M}{\partial K} \,\,,\,\, \frac{\partial K}{\partial \tau}=-\frac{\partial M}{\partial A}\,,
\label{px}
\end{equation} 
where $\tau=1/T$ is the time in Euclidean metric.

These canonically conjugate variables were used in particular for the calculations of the transition rate from the black hole to the white hole with the same mass.\cite{Volovik2022} This rate is determined by the integral $\int_C A(K)dK$ over the tunneling trajectory $C$, which connects black and white holes.
The extended first law of black hole thermodynamics, which includes $K$ as a thermodynamic variable, is $dS=-AdK+(1/T)dM$, see Ref. \cite{Volovik2022} and also Ref. \cite{KimNam2025}.

For the canonically conjugate variables one has the standard uncertainty relation ($\hbar=c=1$):
\begin{equation}
\Delta A \, \Delta K \geq \frac{1}{2}\,,
\label{conjugate}
\end{equation}
The horizon area is $A=\pi M^2/K^2$, and thus for fixed mass $M$ of the black hole, the variation of the gravitational coupling leads to the following variation the horizon area:
 \begin{equation}
\frac{\Delta A}{A}= 2\frac{\Delta K}{K} \,.
\label{ratio}
\end{equation}
Inserting variation $\Delta K=(K/2A) \Delta A$ to Eq.(\ref{conjugate}) one obtains the area fluctuations:
\begin{equation}
\left<( \Delta A)^2\right> \, \geq \frac{1}{K}\left<A\right>= 4G\hbar \left<A\right> \,.
\label{AreaF}
\end{equation}
Here we restored $\hbar$ (keeping $c=1$) to demonstrate that these are quantum fluctuations of the black hole area.

 The coefficient  $4$ in the area variance in Eq.(\ref{AreaF}) differs from the corresponding coefficients  in Refs. \cite{Ciambelli2025,ParikhPereira2024}. In Ref.\cite{Ciambelli2025} the coefficient is $\pi$,
 while in Ref. \cite{ParikhPereira2024} it was concluded that the precise relationship between the computed quantum variance and fluctuations associated with black hole entropy remains unclear. However, the same equation (\ref{AreaF}) can be obtained for the variance of the area of the cosmological horizon. The latter can be obtained by considering thermodynamic fluctuations of the entropy of the de Sitter (dS) spacetime, see the next Section.  

 \section{Thermodynamic fluctuations of de Sitter horizon entropy}
  \label{Thermal}

As distinct from the black hole the dS spacetime is homogeneous and isotropic, and thus can be described by the local thermodynamics.\cite{Volovik2025b} This thermodynamics has the local entropy density:
 \begin{equation}
 s=3KH=3\pi KT \,.
\label{LocalEntropy}
\end{equation}
Here $H$ is the Hubble parameter and $T=H/\pi$ is the local temperature  of the de Sitter state, which in particular determines the ionization rate of an atom in de Sitter environment.
This local temperature is twice the Gibbons-Hawking temperature $T_{\rm GH}= H/2\pi$ associated with the cosmological horizon. Nevertheless, the total entropy $S(V_H)=sV_H$ of the Hubble volume $V_H= (4\pi/3)/H^3$ coincides with the Gibbons-Hawking entropy  of cosmological horizon, $S(V_H)=A/4G$. This demonstrates, that the local entropy obeys the holographic bulk-horizon correspondence.\cite{Hooft1993,Susskind1995} This holographic correspondence connects the extensive entropy of the homogeneous Universe with the non-extensive entropy of the cosmological horizon.

The difference between the local temperature and the Gibbons-Hawking temperature has a natural explanation: these temperatures are measured by two different observers.
The observer measuring the local temperature, such as the activation temperature of the ionization of atoms in de Sitter environment, has all the necessary information. On the other hand, the observer measuring the temperature of the Hawking radiation from the cosmological horizon has no information about the simultaneous creation of the Hawking partner beyond the horizon. This is the reason why this observer underestimates the temperature by a factor of 2. These arguments do not apply to a black hole, where the creation of a Hawking partner is represented by a back reaction.\cite{Wilczek2000}

Since the de Sitter state is homogeneous and isotropic, we can apply the standard equations for the thermodynamic fluctuations (see Sec. 112 in the book \cite{Landau_Lifshitz} by Landau and Lifshitz). Then using the linear dependence of local entropy density on temperature and thus the linear dependence of the heat capacity, one obtains the fluctuations of the local entropy:
 \begin{equation}
\left<( \Delta s)^2\right> = \frac{s}{V} \,.
\label{LocalEntropyF}
\end{equation}
Then for the total entropy $S(V)=sV$ in the arbitrary volume $V$ one has:
 \begin{equation}
\left<( \Delta S)^2\right> = \left<S\right> \,.
\label{EntropyF}
\end{equation}

The Eq.(\ref{EntropyF}) does not contain the Planck constant $\hbar$, which demonstrates that it describes thermodynamic fluctuations. This equation is automatically valid also for the entropy of the Hubble volume with $V=V_H$. Then, since $S(V_H)=sV_H=A/4G\hbar$, 
one obtains for the fluctuations of the area of the cosmological horizon:
\begin{equation}
\left<( \Delta A)^2\right> = 4G\hbar \left<A\right> \,.
\label{AreaFCosm}
\end{equation}
This coincides with Eq.(\ref{AreaF}) for the quantum fluctuations of the area of the black hole horizon. Thus the classical fluctuations of the entropy of the Hubble volume are equivalent to the quantum fluctuations of the area of the cosmological horizon.
 
 The exact relation between the quantum and thermodynamic fluctuations is typical for the systems with horizons. This can be seen also from the processes of splitting of the black hole into the smaller parts. On one hand, such process is quantum, being determined by the macroscopic quantum tunneling, and on the other hand it is determined by the decrease of entropy after splitting,\cite{Wilczek2000,Volovik2022} which is manifestation of the thermal fluctuations. This is one of many examples where gravity serves as a bridge between thermodynamics and quantum mechanics.\cite{Wald1999}
 
Both, the equation (\ref{EntropyF}) and equation (\ref{AreaFCosm}) correspond to the fluctuations of the so-called modular Hamiltonian ${\cal H}$:
 \begin{equation}
\left<( \Delta {\cal H})^2\right> = \left<{\cal H} \right> \,,
\label{ModularF}
\end{equation}
where $\left<{\cal H} \right>=A/4G$, see e.g. Ref. \cite{VerlindeZurek2020} and the references in Ref. \cite{Ciambelli2025}.

 Note that Eq.(\ref{EntropyF}) does not contain the gravitational coupling $K$. This is in agreement with the observation by Jacobson \cite{Jacobson1994} that both $K$ and $S$ are renormalized by the very same quantum fluctuations. 

Finally, note that we consider the ideal de Sitter state, ignoring its instability and decay.\cite{Volovik2024T} The latter disrupt the connection between the properties of the cosmological and black hole horizons, see also Refs.\cite{Trivedi2025a,Trivedi2025b}.

 \section{Black hole fluctuations via ensemble of Planck scale "atoms"}
 \label{Planckons}
 
Another interesting property of the quantum-thermodynamic fluctuations of the black hole is provided by the toy model of the black hole entropy.\cite{Volovik2025P} In this model, the black hole with mass $M$ is represented by the peculiar ensemble of $N$ Planckons --  objects with Planck scale mass.\cite{Markov1967,Hawking1971,Aharonov1987,Rovelli2024,YenChinOng2025,Trivedi2025}. For Planckons with reduced Planck mass $m_{\rm P}$:
\begin{equation}
M=N m_{\rm P} \,\,,\,\,m_{\rm P}^2=\frac{\hbar}{8\pi G}\,.
\label{BWHolesN2}
\end{equation}

The processes of splitting and merging of the black holes demonstrate that the entropy $S_{\rm BH}(N)$ of the black hole is not determined by the individual Planckons, which have zero entropy, $S(N=1)=0$. The entropy of black hole is determined by the correlated (or entangled) pairs of Planckons. Each pair has entropy $S(N=2)=1$, and the entropy of the black hole is determined by the number ${\cal N}$ of pairs:
\begin{equation}
S_{\rm BH}(N)={\cal N} =\begin{pmatrix}N
\\ 2\end{pmatrix} =\frac{N! }{2! (N-2)!}=\frac{N(N-1)}{2}  \,.
\label{BWHolesN3}
\end{equation}
 
 In the thermodynamic limit $N\gg 1$, which according to Eq.(\ref{BWHolesN2}) also corresponds to the classical limit $\hbar \rightarrow 0$, the equation (\ref{EntropyF}) gives for fluctuations of the Planckon number:
 \begin{equation}
\left<( \Delta N)^2\right> = \frac{1}{2}  \,.
\label{PlanckonF}
\end{equation}
 Such unusual variance of the number of Planckons $N$, which is not proportional to $N$, is the consequence of the Tsallis-Cirto non-extensive statistics\cite{TsallisCirto2013,Tsallis2020} with $\delta=2$. This statistics describes the non-extensive composition law for entropy in the processes of splitting and merging of black holes (see Refs.\cite{Volovik2025TC,Manoharan2025} and references therein):
\begin{equation}
\sqrt{S(N_1 +N_2)}= \sqrt{S(N_1)} +\sqrt{S(N_2)}\,.
\label{BWHolesN}
\end{equation}
This non-extensive composition law is in the origin of the black-hole thermodynamics, which distinguishes the Tsallis-Cirto  $\delta=2$ entropy from other possible types of generalized entropy.\cite{Odintsov2022a,Odintsov2022b,Odintsov2025} 
The same composition law for the entropies of the inner and outer horizons of the rotating Kerr black hole gives the following entropy of the Kerr black hole:\cite{Volovik2025Kerr} 
\begin{equation}
S_{\rm Kerr}(N,J)=4\pi N(N-1) + 4\pi\sqrt{J(J+1)}\,,
\label{Kerr}
\end{equation}
where the Plackons with Planck mass $M_{\rm P}=\sqrt{\hbar/G}$ were used.

 On the other hand, the ensemble of the ${\cal N}$ pairs of Planckons has conventional extensive composition law:
 \begin{equation}
 S({\cal N}_1 +{\cal N}_2)=S({\cal N}_1) +S({\cal N}_2) \,,
\label{extensive}
\end{equation}
and the conventional fluctuations, which are consistent with the Poisson distribution:
\begin{equation}
\left<( \Delta {\cal N})^2\right> =  {\cal N} \,.
\label{PlanckonpairsF}
\end{equation}
This is the representation of the equations (\ref{AreaF}), (\ref{EntropyF}) and (\ref{ModularF}) for quantum and thermodynamic fluctuations in integer numbers. In principle, this may suggest the realization of the Bekenstein idea of quantization of the horizon entropy.\cite{Bekenstein1974} 

The equation (\ref{PlanckonF}) also suggests the following variance of the black hole mass $M$:
 \begin{equation}
 \left<( \Delta M)^2\right> = \frac{1}{2}m_{\rm P}^2 \,.
\label{MassF}
\end{equation}
This shows that, unlike the traditional approach (see e.g. Ref. \cite{Avramov2024} and references therein), in the toy model the black hole is robust to thermodynamic fluctuations. This is because in the toy model the black hole is the equilibrium state. The Hawking radiation represents the thermal/quantum fluctuation. After the emission of Planckons, the entropy of the black hole decreases, but it is subsequently restored in the natural process of absorption of the emitted Planckons.

 \section{Discussion}
 \label{discussion}
 
  The event horizon is the source of irreversibility, which is similar to the statistical irreversibility.\cite{Witten2025} That is why for systems with horizon, there is no difference between the quantum and thermal fluctuations. The quantum processes of quantum tunneling determine the thermodynamics of these systems, with their temperatures, entropies and fluctuations. 
  We considered three examples of the entropy variance, which support this view.
  
  (i) The variance of the area of the black hole horizon can be obtained in the domain of quantum fluctuations. Here we used the uncertainty principle, which relates the variances of the canonically conjugate variables: the gravitational coupling $K=1/4G$ and black hole area $A$. This gives Eq.(\ref{AreaF}) for the variance of the horizon area.
  
   (ii) The variance of the entropy of the cosmological horizon in de Sitter state can be obtained in the domain of classical thermodynamics fluctuations. The entropy of the cosmological horizon is equal to the entropy of the Hubble volume, thus representing the holographic bulk-horizon connection. As a result, the variance of the entropy can be obtained from the conventional thermodynamics in bulk, which determines thermal fluctuations of thermodynamic variables, such as entropy. The variance of the entropy in Eq.(\ref{EntropyF}), which is obtained in the domain of thermal fluctuations, is applicable to the arbitrary volume $V$ of the homogeneous de Sitter state. When applied to the Hubble volume $V=V_H$, one obtains the variance of the cosmological horizon area in Eq.(\ref{AreaFCosm}) in agreement with the equation (\ref{AreaF}), which was obtained for the black hole horizon in the domain of quantum fluctuations.
   
   (iii) The variance of entropy can be expressed in terms of integer numbers $N$ in the Planckon model, which simulates the Tsallis-Cirto non-extensive ensemble with $\delta =2$. The variance is determined by the Poisson distribution of the correlated pairs of Planck scale objects -- Planckons. For large number of Planckons the result agrees with the results obtained in (i) and (ii). The Planckon model simulates the Bekenstein-type quantization of the horizon entropy.

\end{document}